%% file: main.tex
\documentclass[10pt, conference, letterpaper]{IEEEtran}
\pdfoutput=1 
\usepackage{booktabs} 
\usepackage{color,fancyvrb}
\usepackage{subfig}
\usepackage[normalem]{ulem}

\usepackage{comment} 
\usepackage{amsmath}
\usepackage[T1]{fontenc}
\usepackage{tabularx}
\usepackage{tabu}
\usepackage{multirow}
\usepackage[bookmarks=false]{hyperref}
\usepackage{xcolor}
\usepackage{listings}
\usepackage{caption}
\usepackage{tikz}
\usepackage{lipsum}
\newcommand{\commentBy}[3]{\textcolor{#1}{\textbf{#2:} #3}}

\newif\ifcommentson

\newcommand{\ste}[1]{\ifcommentson \commentBy{magenta}{SS}{#1} \fi}

\newif\ifextended
\newif\ifshortver

\extendedtrue

\newcommand{\extended}[1]{\ifextended \ifshortver \textcolor{purple}{#1} \else \textcolor{black}{#1} \fi  \fi}
\newcommand{\shortver}[1]{\ifshortver \ifextended \textcolor{blue}{#1} \else \textcolor{black}{#1} \fi \fi}

\begin{document}
\begin{NoHyper}

\title{An Efficient Linux Kernel Implementation of Service Function Chaining for legacy VNFs based on IPv6 Segment Routing}




%
%

\IEEEtitleabstractindextext{
\begin{abstract}
We consider the IPv6 Segment Routing (SRv6) technology for Service Function Chaining of Virtual Network Functions (VNFs). Most of the VNFs are \textit{legacy VNFs} (not aware of the SRv6 technology) and expect to process \textit{traditional} IP packets. An \textit{SR proxy} is needed to support them. We have extended the implementation of SRv6 in the Linux kernel, realizing an open source SR-proxy, referred to as SRNK (SR-Proxy Native Kernel). The performance of the proposed solution (SRNKv1) has been evaluated, identifying a poor scalability with respect to the number of VNFs to be supported in a node. Therefore we provided a second design (SRNKv2), enhancing the Linux Policy Routing framework. The performance of SRNKv2 is independent from the number of supported VNFs in a node. We compared the performance of SRNKv2 with a reference scenario not performing the encapsulation and decapsulation operation and demonstrated that the overhead of SRNKv2 is very small, on the order of 3.5\%.\\
\end{abstract}

}


\author{\IEEEauthorblockN{
Andrea Mayer\IEEEauthorrefmark{1}\IEEEauthorrefmark{2},
Stefano Salsano\IEEEauthorrefmark{1}\IEEEauthorrefmark{2},
Pier Luigi Ventre\IEEEauthorrefmark{1},
\\Ahmed Abdelsalam\IEEEauthorrefmark{3}\IEEEauthorrefmark{4},
Luca Chiaraviglio\IEEEauthorrefmark{1}\IEEEauthorrefmark{2},
Clarence Filsfils\IEEEauthorrefmark{4}
}
\IEEEauthorblockA{
\IEEEauthorrefmark{1}University of Rome Tor Vergata,
\IEEEauthorrefmark{2}CNIT,
\IEEEauthorrefmark{3}Gran Sasso Science Institute,
\IEEEauthorrefmark{4}Cisco Systems
}
\\ 
\extended{\textbf{Extended version of the paper accepted to IEEE Netsoft 2019 - v04 - July 2019}}
}

\maketitle

\IEEEdisplaynontitleabstractindextext
\IEEEpeerreviewmaketitle

\input{inc/01-introduction}

\input{inc/02-sfc}

\input{inc/03-sr-design}

\input{inc/04-test-environm}

\input{inc/05-performance-analysis}
\input{inc/06-related-work}
\input{inc/07-conclusions}
\section*{Acknowledgment}
    This work has received funding from the Cisco University Research Program.

\bibliographystyle{IEEEtran}
\bibliography{main}

\extended{
\newpage
\input{inc/appendix_a}
\input{inc/appendix_b}
\input{inc/extended-results.tex}
\input{inc/extended-nsh.tex}
}

\end{NoHyper}
\end{document}

%% file: inc/01-introduction.tex
\section{Introduction}
\label{sec:introduction}

\ste{We will have an extended version on arxiv. \cite{legacy-vnf-sr-proxy-linux-kernel-extended}\\The text that will only be in the extended version is marked with the command \textbackslash extended\{...\}\\The text that will only be in the netsoft version is marked with the command \textbackslash shortver\{...\} }

Network Operators are facing difficult challenges to keep up with the increasing demand for capacity, the need to support fast service creation and at the same time the goal of reducing the costs. Network Function Virtualization (NFV) \cite{nfv} \cite{mano} and Software Defined Networking (SDN) represent an answer to these challenges and are changing the way IP networks are designed and operated. Leveraging Cloud Computing principles, NFV moves the traditional data-plane network functions from expensive, closed and proprietary hardware to the so-called Virtual Network Functions (VNFs) running over a distributed, cloud-like infrastructure referred to as NFVI (NFV Infrastructure). The SDN architecture splits the data and control planes and moves the intelligence to the SDN controller. SDN aims at simplifying the introduction of new services and fostering flexibility thanks to the centralized network state view.

The concept of services chaining (also known as Service Function Chaining - SFC \cite{RFC7665}) is directly associated to NFV. Actually, the idea of creating a processing path across services pre-dates the NFV concept as stated in \cite{RFC7498} and \cite{sfc_wh}. In fact, service chaining has been traditionally realized in a static way by putting hardware functions as middle-points of the processing paths and in some cases by diverting the forwarding paths with manual configuration of VLANs stitching or policy routing. However, these ``static'' approaches comes with several drawbacks which are detailed in \cite{RFC7498}. In particular, they are intrinsically difficult to scale and hard to reconfigure. On the other hand, the current view of SFC applied to NFV is that it has to be highly \textit{dynamic} and \textit{scalable}.\extended{The IETF SFC Working Group (WG) has investigated the scenarios and issues related to dynamic service chaining \cite{RFC7498} and proposed a reference architecture \cite{RFC7665}. The main logical elements of this architecture are i) Classifiers; ii) Service Functions Forwarders (SFF), iii) the Service Functions, iv) SFC proxies. The Classifiers match the traffic against a set of policies in order to associate the proper Service Function Chain. The SFFs forward the traffic towards the Service Functions or towards other SFFs and handle the traffic coming back from the Service Functions. The SFC framework proposed in \cite{RFC7665} does not pose any restriction on the function that can be chained: they can be both virtualized (VNFs) or physical functions (PFs). For the sake of simplicity, hereafter in the paper we will only refer to the virtualized case (which we believe is the most significant) and will simply use the term VNF instead of Service Function. In this scenario, the forwarding of traffic along a Service Chain needs to be supported by specific protocols and mechanisms that allow the architectural elements to exchange context information. The VNFs can participate to these chaining mechanisms and in this case they are called \textit{SFC aware}. On the other hand, the legacy VNFs that do not interact with the SFC protocols and mechanisms are called \textit{SFC unaware}. The SFC proxy elements are needed for the latter type of VNFs. An SFC proxy hides the SFC mechanisms to the SFC unaware VNFs, that will receive and send plain IP traffic. 
The IETF SFC WG is considering the Network Service Header (NSH) \cite{nsh-id} as a specific solution for the realization of the SFC architecture. The NSH header defines the service-level data-plane encapsulation for realizing the VNFs chaining. The NSH header identifies a service chain which is associated to the packets. Moreover, it defines the packet meta-data that can be inserted into the header to exchange state between the nodes of the SFC architecture.}
In this work we are advocating the use of IPv6 Segment Routing (SRv6) to implement Service Function Chaining \cite{srv6-sfc, lebrun2015leveraging}. Segment Routing \cite{filsfils2015segment, idsrarch} is a form of source routing, which allows to add a sequence of \textit{segments} in the packet headers to influence the packet forwarding and processing within the network. Segment Routing has been designed and implemented for the MPLS and IPv6 data planes, we only focus here on the IPv6 version, denoted as \textit{SRv6}.  In the SRv6 architecture, the \textit{segments} are expressed as IPv6 addresses. The SRv6 network programming model \cite{srv6-net-prog}, leveraging the huge IPv6 addressing space, extends the SRv6 architecture from a simple forwarding mechanism for steering packets to a more general network programming abstraction. A segment can represent an \textit{instruction} or \textit{behavior} and not only a \textit{network location}. Our proposed approach is fully aligned with the network programming model described in \cite{srv6-net-prog}. The SRv6 architecture is not limited to Service Function Chaining, which represents only a possible use case. Indeed, SRv6 can support several applications in a network provider backbone like Traffic Engineering, Network Resilience (Fast Rerouting), Virtual Private Networks (VPNs), Multicast, Content Delivery Networks (CDNs). \extended{With respect to the MPLS based data plane, SRv6 it has the advantage that it can be better integrated in host networking stack. For this reason Data Center applications could also benefit from SRv6.}

A relevant subset of the SRv6 \cite{ID-ipv6-SRH} and network programming model \cite{srv6-net-prog} specifications have been implemented and integrated in the mainline Linux kernel \cite{lebrun2017implementing}. In this paper, we rely on this existing work and extend it to focus on the Service Function Chaining of legacy VNFs, which are not able to process the SRv6 headers. The support of legacy VNFs is important for Internet Service Providers (ISP) for different reasons: i) it guarantees a feasible migration strategy saving past investments; ii) it facilitates the interoperability and the multi-vendor scenarios, i.e deployments composed by VNFs of different vendors; iii) the development of SRv6 aware VNFs requires a new implementation cycle which can be more expensive in the short period.

\begin{figure}
    \centering
    \includegraphics[width=0.33\textwidth]{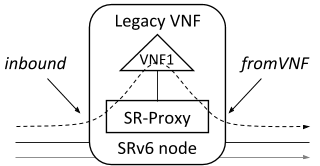}
    \caption{SRv6 NFV Node with SR-Proxy for SR-unaware VNF}
    \label{fig:srv6_nfv}
    \vspace{-3ex}
\end{figure}

As introduced above, a proxy element needs to be inserted in the processing chain as relay mechanism in order to support SRv6 unaware VNFs (see Figure~\ref{fig:srv6_nfv}). The latest Linux kernel still lacks of the functionality to implement such SRv6 proxy element. In a prior work (SREXT \cite{vnf-chain-srv6}), we have provided this functionality as an external module not integrated with the most recent SRv6 developments in the Linux kernel. Considering the importance of the support of legacy SR-unaware applications in NFV deployments, the main contribution this paper is the design and implementation of an \textit{SR-proxy} integrated in the Linux kernel networking components. We refer to this work as SRNK (SR-Proxy Native Kernel). We designed a first version of SRNK and evaluated its performance, identifying a poor scalability with respect to the number of VNFs to be supported. The issue is actually related to the implementation of Policy Routing framework in Linux. Therefore we provided a second design, enhancing the Linux Policy Routing framework, whose performance does not depend on the number of supported VNFs in a node.  

The content of the paper is as follows. Section \ref{sec:sr_sfc} introduces SFC based on SRv6 considering both SRv6 aware and unaware VNFs. The proposed design and implementation of SRv6 Proxy to support legacy VNFs in the Linux kernel is described in Section \ref{sec:sr_proxy}. Our testing environment and methodologies for performance analysis are reported in Section \ref{sec:sr_test}. Sections \ref{sec:sr_perf} details the performance evaluation of the implemented solutions. Finally, in Section~\ref{sec:conclusions} we draw some conclusions and discuss future work.

This work has been performed in the context of the ROSE research project \cite{rose} which focuses on the development of an open source SRv6 ecosystem. The source code of all components of SRNK including the patches to the user space utilities are freely available at \cite{srnk-home-page}. 

%% file: inc/02-sfc.tex
\section{SFC Based on IPv6 Segment Routing}
\label{sec:sr_sfc}

The Segment Routing architecture is based on the \textit{source routing} approach (\cite{idsrarch} and \cite{filsfils2015segment}): it is possible to include a list of instructions (the so called \textit{segments}) in the packet headers. \extended{A comprehensive survey on Segment Routing can be found in \cite{sr-survey}}

This work considers the use of SRv6 for SFC, leveraging its scalability properties.\extended{Thanks to the source routing approach, SRv6 is able to simplify network operations.} Generally speaking, the advantage of approaches based on source routing lies in the possibility to add state information in the packet headers, thus avoiding or minimizing the information that needs to be configured and maintained by the internal nodes. The possibility to interact only with the edge nodes to setup complex services \shortver{allows simpler and faster service setup and re-configuration.}\extended{is extremely appealing from the point of view of simplicity and efficiency. This greatly improves the scalability of services based on SR and allows simpler and faster service setup and re-configuration. In \cite{interconnecting} the scaling capability of Segment Routing has been demonstrated considering an use case of 600,000 nodes and 300 millions of endpoints.}

By exploiting the SRv6 approach the VNFs can be mapped in IPv6 addresses in the \textit{segments} list (SIDs list in SRv6 jargon) and we can represent the VNF chain using this list carried in the Segment Routing Header (SRH).

The SR information can be pushed into the packets using two different approaches, denoted as \textit{insert} and \textit{encap} modes, respectively.\extended{According to the SRv6 network programming document \cite{srv6-net-prog}, when a node uses the \textit{insert} mode the SRH is pushed as next header in the original IPv6 packet, immediately after the IPv6 header and before the transport header. The original IPv6 header is changed, in particular the \textit{next header} is modified according to the value of SRH, the IPv6 destination address is replaced with the IPv6 address of the first SID in the segment list, while the original IPv6 destination address is carried in the SRH header as the last segment of the list.} In this work we only consider the \textit{encap} mode: the original IPv6 packet is transported as the inner packet of an IPv6-in-IPv6 encapsulated packet and travels unmodified in the network. The outer IPv6 packet carries the SRH header with the segments list.\extended{\footnote{As any tunneling (encapsulation) method, SRv6 introduces overhead the packets. The insert mode introduces an overhead of $8 + N * 16 [bytes]$ where $N$ is the number of segments, while in the \textit{encap} mode the overhead is $40 + 8 + N * 16 [bytes]$.}}

An \textit{SR-aware} VNF can process the SRH of the incoming packets and can use it to influence the processing/forwarding of the packets. Such VNFs interact with the node Operating System or with SR modules in order to read and/or set the information contained in the SRH. On the other side, the \textit{SR-unaware} (legacy) VNFs are not capable to process the SRv6 SFC encapsulation. In this scenario an \textit{SR proxy} is necessary to remove the SRv6 header and deliver a ``clean'' IP packet to the VNF. Figure~\ref{fig:srv6_nfv} provides the reference architecture for a SRv6 NFV node that includes an \textit{SR-unaware} VNF (VNF1 in the Figure). We refer to packets incoming to the SRv6 NFV node that should be forwarded to the VNF by the SR-proxy as \textit{inbound} packets. The SR-Proxy needs to intercept the packets coming out from the VNF and re-apply the SRv6 SFC encapsulation. We refer to these packets as \textit{fromVNF} packets.

In \cite{srv6-sfc}, a set of SR-proxy \textit{behaviors} have been defined, among them we mention: i) \textit{static} proxy (also called End.AS behavior); ii) \textit{dynamic} proxy (End.AD behavior); iii) \textit{masquerading} proxy (End.AM behavior). The first two cases (\textit{static} and \textit{dynamic} proxies) support IPv6 SR packets in \textit{encap} mode. The encapsulated packets can be IPv6, IPv4 or L2 packets. The SR proxy intercepts SR packets before being handed to the \textit{SR-unaware} VNF, hence it can remove the SR encapsulation from packets. For packets coming back from \textit{SR-unaware} VNF, the SR proxy can restore the SRv6 encapsulation updating the SRH properly. The difference between the \textit{static} and the \textit{dynamic} proxies is that the SR information that needs to be pushed back in the packets is statically configured in the first case and it is \textit{learned} from the incoming packets in the \textit{dynamic} case.\extended{Instead, the \textit{masquerading} proxy supports SR packets travelling in insert mode. It \textit{masquerades} the SR packets before they are sent to the legacy VNF by replacing the IPv6 destination address (the current SID of the segment list) with the original IPv6 destination (i.e. the last segment in the SID list).}\extended{It is assumed that a VNF compatible with this operating mode is processing IPv6 packets and does not alter the SRH, it just ignores it. In this way, when packets are received back, the SR proxy can restore the correct information in the IPv6 header in a stateless way, just using the information contained in the SRH.}

\begin{figure}
    \centering
    \includegraphics[width=0.48\textwidth]{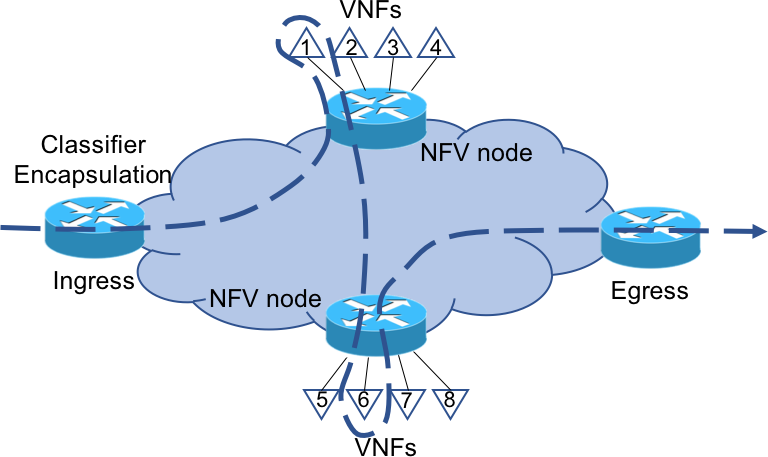}
    \caption{SFC scenario}
    \label{fig:sfc_scenario}
    \vspace{-3ex}
\end{figure}

Let us discuss the operational model and the state information that need to be configured and maintained in the SRv6 NFV nodes. Figure~\ref{fig:sfc_scenario} illustrates a SRv6 based NFV domain, in which the VNFs are hosted in different NFV nodes. The packets to be associated to VNF chains are classified in \textit{ingress} nodes, where the SR encapsulation is added. A network operator willing to use SRv6 SFC chaining for \textit{SR-unaware} VNFs, will first need to associate VNFs to Segment IDs (SIDs) in the hosting SRv6 NFV nodes. We recall that a SID is represented by an IPv6 address. Each SRv6 NFV node has a pool of IPv6 addresses (prefixes) that are available to be used as SIDs for its VNFs. These prefixes are distributed using regular routing protocols, so that the reachability of all VNFs is assured. The association of the IPv6 address SID to a VNF is a configuration operation to be performed in the SRv6 NFV node and it binds the SID to the virtual interface that connects the SR-proxy to the VNF. This operation is performed when a legacy VNF is created in a NFV node. The corresponding state information is used in the \textit{inbound} direction, when packets directed to the VNF are processed by the SR-proxy. The second step is to configure a VNF chain across the VNFs that are running over the SRv6 NFV nodes. The VNF chain will be applied to a packet by inserting a SID list in the IPv6 SR header in the ingress node. Therefore, the classification of packets and the association with the SID list has to be configured in the ingress node. Each NFV node which runs a legacy VNF needs the proper information to process the packets in the \textit{fromVNF} direction. This is done differently for the respective types of proxy. \shortver{For example, the dynamic proxy ``learns'' the SID list from packets in the \textit{inbound} direction and saves it as state information. Thanks to this information, the proxy re-associates a SID list to a packet coming from a specific VNF. Therefore this solution works under the assumption that a VNF in a node is associated to a single chain. If the same VNF type needs to be associated to different chains in the same node, additional separate instances of the VNF are needed. This could not be a problem if lightweight virtualization mechanisms like Containers or even Unikernels are used. In any case, it is the price to be paid for not performing a classification based on packet inspection for the packets coming from the VNF, as needed in traditional SFC solutions.} \extended{In the \textit{static} proxy case (End.AS behavior), the state information needed to process the packets coming from the VNF is done by statically configuring the SR-proxy with the SID list to be re-inserted in the packet. Both the \textit{dynamic} proxy (End.AD behavior) and the \textit{masquerading} one (End.AM behavior) have the good property that they do not need to be configured when a new chain is added. The dynamic proxy ``learns'' the SID list from the packets in the \textit{inbound} direction (and so it saves a state information). The \textit{masquerading} proxy does not even need to save the state information as the SID list is carried along with the packet through the legacy VNF (which has to be IPv6 and needs to accept the SRH header without interfering with it). Table \ref{table:sr-proxy-comp} compares the different SR proxy behaviors.
\begin{table}[ht]
\caption{Comparison of SR proxy behaviours}
\centering 
\begin{tabular}{ |m{2.4cm} |m{1.74cm} |m{1.74cm} |m{1.1cm}| } 
\hline
& End.AD & End.AS & End.AM \\ 
\hline
Generate traffic & Yes & Yes & No \\ 
\hline
Modify packets   & Yes & Yes & No \\ 
\hline
Stateless        & No  & No & Yes\\ 
\hline
State-config    & Auto  & Manual & N/A \\ 
\hline
Traffic supported  & IPv4/IPv6/L2 & IPv4/IPv6/L2 & IPv6 \\ 
\hline
\end{tabular}
\label{table:sr-proxy-comp} 
\end{table}
}
In this work, we focus on the design and in-kernel implementation of the SR \textit{dynamic} proxy as it represents the most versatile solution (being able to support legacy VNFs working with IPv6, IPv4 and L2 packet) and it offers a simple operational model.

%% file: inc/03-sr-design.tex
\section{Design of the SRv6 Proxy}
\label{sec:sr_proxy}

This section describes the design and implementation aspects of the SR-proxy. We start with subsection~\ref{sec:general} that provides a brief introduction of general concepts that will be extensively used in the paper. It briefly describes the network programming model defined in \cite{srv6-net-prog} and how it has been implemented inside the Linux kernel. Then, it presents the so called policy routing that introduces a match-action framework for IP routing in Linux and finally it shows our previous SREXT solution. Subsection~\ref{sec:proxy1} presents the first implementation of our \textit{SR-Proxy} integrated in the kernel, referred to as SRNKv1 (Native kernel v1) and elaborates on its operations. In subsection~\ref{sec:proxy2} we analyze the performance issues of SRNKv1 and present our second design and implementation (SRNKv2), discussing its performance improvements.

\subsection{General Concepts and State-of-the-art}
\label{sec:general}

\subsubsection{Network Programming Model}

The SRv6 network programming model \cite{srv6-net-prog} extends the IPv6 Segment Routing concept from the simple steering of packets across nodes to a general network programming approach. Quoting from \cite{srv6-net-prog} ``Each segment represents a function to be called at a specific location in the network'', a function can span from a simple action like forwarding or a complex processing defined by the user. Going into the details, each SRv6 capable node maintains the so-called \textit{My Local SID Table} \cite{srv6-net-prog}, each entry of this table maps a \textit{segment} (SID) into a local function. As a consequence, when a packet enters in an SRv6 enabled node with an active \textit{segment} matching an entry of the table, the associated function is applied to the packet. Leveraging the fact the \textit{segments} are represented as regular IPv6 addresses, the node can advertise them using any routing protocol. Combining these ``network instructions'' it is possible to literally program the network and realize very complex network behaviors.

The association of a function to a SID resembles the execution of the nexthop lookup function in the IP nodes. Indeed, \textit{My Local SID Table} has been realized in the Linux networking stack (from kernel 4.14) using an IPv6 routing table that contains routes on which custom processing function are associated. In recent Linux kernel implementations, \textit{lightweight tunnel} (LWT) provides the capability of performing a generic operation on the packets (which can span from a simple encap/decap to a general purpose processing). The Linux SRv6 network programming implementation leverages the mechanism offered by the Linux kernel that allows to associate LWTs to routing entries.

\begin{figure}
    \centering
    \includegraphics[width=0.44\textwidth]{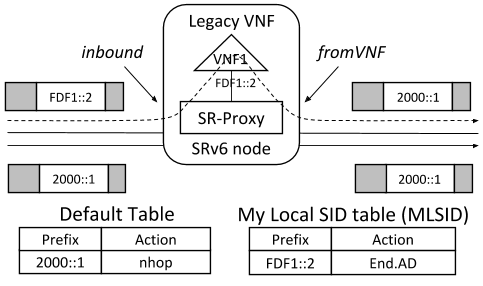}
    \caption{SRv6 node processing}
    \label{fig:srv6_processing}
    \vspace{-3ex}
\end{figure}

The \texttt{seg6local} LWT is the specific type of lightweight tunnel that supports the SRv6 network programming features in the Linux kernel \cite{lebrun2016design}. Starting from Linux kernel 4.14 a subset of the behaviors described in \cite{srv6-net-prog} have been implemented, while the SR proxy behaviors are not supported yet. The purpose of this work is to extend the implementation of the SRv6 network programming model currently available in the Linux kernel to support the \textit{dynamic} proxy (End.AD behaviour). Figure~\ref{fig:srv6_processing} shows the processing of a SRv6 node where a legacy VNF is deployed. 

Let us refer to Figure~\ref{fig:srv6_processing} to explain the details of SRv6 processing in an NFV node hosting an SR-proxy. For the packets in the \textit{inbound} direction the SR-proxy classifies the packets based on the IPv6 destination address, decapsulates them as needed and forwards to the proper interface towards the VNF. For the packets in the \textit{fromVNF} direction (i.e. sent back by the \textit{SR-unaware} applications), the SR-proxy needs to restore the SRH header after the identification of the interface from where the packets are coming. 
Looking at Figure~\ref{fig:srv6_processing} an \textit{inbound} packet having as destination address which does not correspond to a VNF (e.g. \texttt{2000::1}) are simply forwarded by the node over an outgoing interface (\texttt{oif}), looking at the default routing table. The packets having \texttt{FDF1::2} as IPv6 destination address (and active \textit{segment} in the segment list) is matched by the node in \textit{My Local SID Table}, hence the \textit{SR-proxy} behavior is applied and the packet is forwarded to the VNF1. When considering the packets coming from the legacy VNF1, the proxy restores correctly the SRv6 header and delivers it to the IPv6 processing of the node that will forward to the next hop.
Note that \textit{My Local SID Table} and the normal routing table does not need to be separated, this is actually an implementation aspect. In the current Linux implementation the SID entries can be inserted in any routing table, therefore also in the default routing table.

\subsubsection{Policy Routing}
\label{sec:policy_routing}

Policy Routing extends the traditional routing based only on IP destination addresses. With Policy Routing the forwarding decision on a packet can be based on different features of the packets, considering packet headers at different protocol levels, incoming interfaces, packet sizes and so on. According to this extended routing model, the Linux kernel implementation of Policy Routing complements the conventional destination based routing table (that leverages the longest prefix match) with a Routing Policy DataBase (RPDB). In general, each \textit{Policy Routing} entry in the RPDB consists of a selector and an action. The rules within the RPDB are scanned in decreasing order of priority. If the packet matches the selector of an entry the associated action is performed, for example an action can direct the packet to a specific routing table. The most important consideration for our purposes is that the RPDB rules are sequentially processed, so the performance penalty of checking the rules increases linearly with the number of rules.  

\subsubsection{State-of-the-art - SREXT module}
\label{sec:srext}

The SREXT module (\cite{vnf-chain-srv6}) is our first implementation of the SRv6 network programming model. When it was designed, the Linux kernel only offered the basic SRv6 processing (End behavior). SREXT is an external module that complemented the SRv6 Linux kernel implementation providing a set of behaviors that were not supported yet. Currently most of the behaviors implemented in SREXT are supported by the mainline of Linux kernel (with the exception of the SR-proxy behaviors). So, following this trend, we decided to implement SR-proxy behaviors that were only available using SREXT directly into the kernel avoiding any extra module functionality and dependency.
\shortver{In section~\ref{sec:rel_work} we discuss the shortcomings of SREXT compared to SRNK. More info on SREXT can be found in  \cite{legacy-vnf-sr-proxy-linux-kernel-extended}.}
\extended{In the related work (section~\ref{sec:rel_work}) we analyze the SREXT shortcomings compared to our solution. As for the SR-proxy, SREXT handles the processing of SR information on behalf of the SR-unaware VNFs, which are attached using two interfaces. SREXT provides an additional local SID table which coexists with the one maintained by the Linux kernel. The SREXT module registers itself as a callback function in the \textit{pre-routing} hook of the netfilter \cite{netfilter} framework. Since its position is at beginning of the netfilter processing, it is invoked for each received IPv6 packet. If the destination IPv6 address matches an entry in the local SID table, the associated behavior is applied otherwise the packet will follow the normal processing of the routing subsystem.
\\
A secondary table (the so called ``srdev'' table) is used by SREXT for correctly executing the processing of the \textit{inbound} and \textit{fromVNF} packets. As regards the former, once the packet has passed the sanity check and the SRv6 behavior has been applied, SREXT stores in this table the \textit{fromVNF} interface (where SREXT will receive back the packet from the VNF), the applied behavior, the original IPv6 header and its SRH.  On the \textit{fromVNF} side, the receiving interface is used as a look-up key in the table ``srdev'', if an entry is found the headers are re-added (IPv6 control traffic like NDP is dropped) and finally the packet will go through the kernel IP routing sub-system for further processing. A new \textit{cli} has been implemented for controlling SREXT behaviors and showing its tables and counters.
}

\subsection{SRNKv1}
\label{sec:proxy1}


In this section we present the design of our first kernel implementation of the \textit{dynamic} proxy (End.AD behavior), referred to as \textit{SRNKv1}.
\extended{Most of the following design choices apply also to the static proxy (End.AS behavior), which can be seen as a by-product of the the \textit{dynamic} proxy implementation. In order to simplify the discussion we just mention the \textit{dynamic} proxy in the paragraphs and in the images.}
\textit{SRNKv1} design relies on two distinct LWTs which manage respectively the \textit{inbound} and \textit{fromVNF} traffic. For each LWT, state information is maintained in order to correctly perform the proxy operations. In particular, the \textit{inbound} processing needs an entry on the \textit{My Local SID Table} and uses a per-network namespace hashtable \textit{(per-netns hashtable)} to store the headers that have to be restored during the \textit{fromVNF} processing.

As regards the traffic coming from the legacy VNF, a policy routing entry for each VNF is necessary to classify the packets, a routing table with a default route pointing to the LWT is used for the VNF and finally the \textit{per-netns hashtable} is used to read the headers stored previously by the \textit{inbound} processing. Figures~\ref{fig:proxy1} show an high-level view of the processing inside a SRv6 enabled node and how IPv6 routing network subsystem interacts with the SRv6 \textit{dynamic} proxy implementation.

\subsubsection{Inbound processing}
\label{sec:inbound_processing}

\begin{figure*}[h!tb] 
  	\centering
  	\subfloat[Inbound processing]{%
       \includegraphics[width=0.48\textwidth]{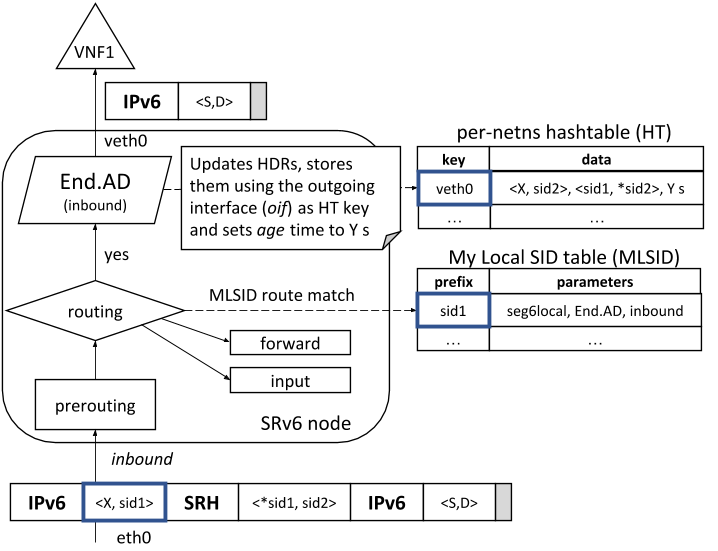}
    \label{fig:proxy1_inbound}}\hfill
  	\subfloat[FromVNF processing]{%
  		\includegraphics[width=0.50\textwidth]{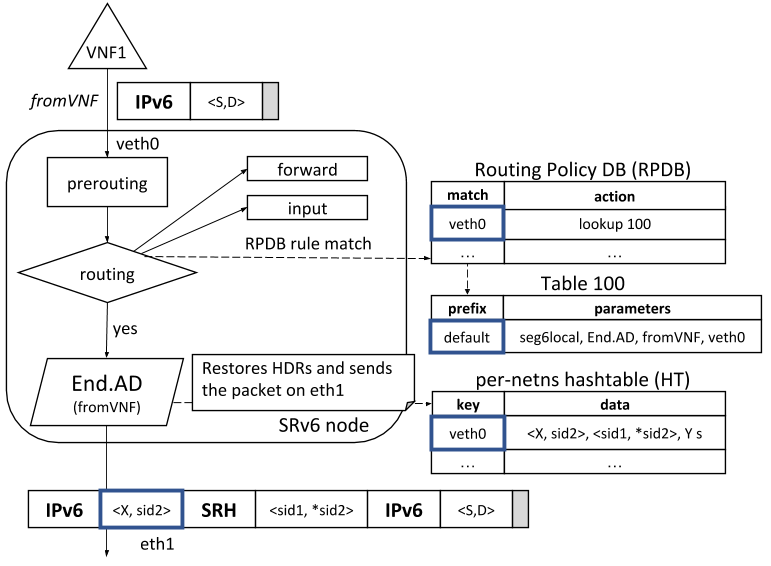}
    \label{fig:proxy1_fromvnf}}
	\caption{SRNKv1 design}
  	\label{fig:proxy1}
  	\vspace{-3ex}
\end{figure*}

\shortver{The \textit{inbound} processing is depicted in Figure~\ref{fig:proxy1_inbound}. During the \textit{routing} phase, if the destination IPv6 address matches the \texttt{sid1} prefix stored in \textit{My Local SID Table}, the Linux kernel executes the \extended{\textit{dynamic} proxy} processing function associated with the LWT route (denoted as End.AD): i) it pops the outer IPv6 and SRv6 headers from the incoming packet; ii) it updates the SID pointer of the SRv6 header to select the next one; iii) it stores such retrieved headers into a \textit{per-netns  hashtable} data structure; iv) it sends out the decapsulated IPv6 plain packet to its designated legacy VNF.}
\extended{The \textit{inbound} processing is depicted in Figure~\ref{fig:proxy1_inbound}. As soon as an IPv6 packet arrives at interface \texttt{eth0} of the NFV node, it enters into the Linux networking stack. After passing the \texttt{pre-routing} stage, the kernel tries to look up the route with the longest prefix that matches the active \textit{segment} of the packet. Due to policy-routing settings, the Linux kernel looks first at \textit{My Local SID Table} and if no matching route has been found, it considers the other tables and possibly moves on the next stages of the processing (\texttt{input} or \texttt{forward}). Figure~\ref{fig:proxy1_inbound} shows this process in details, the packet destination address matches with prefix \texttt{sid1} and the correspondent route is used. Therefore, the Linux kernel executes the processing function associated with the route: the \textit{inbound} End.AD operation. The inbound End.AD operates in three different stages: i) it pops the outer IPv6 and SRv6 headers from the incoming packet; ii) it updates the SID pointer of the SRv6 header to select the next one; iii) it stores such retrieved headers into a \textit{per-netns hashtable} data structure; iv) it sends out the decapsulated IPv6 plain packet to its designated legacy VNF.
\\
Removed headers at step (i) are indexed in the \textit{per-netns hashtable} by using the identifier of the packet outgoing interface (\texttt{oif}), the one used to communicate with the legacy VNF (\texttt{veth0} in Figure~\ref{fig:proxy1_inbound}). Due to the necessity of sharing IPv6 and SRv6 headers between \textit{inbound} and \textit{fromVNF} processing, the choice of storing them within a external shared data structure turned out to be the right solution. This design simplifies the access pattern to the stored data, as well as it increases performance. Indeed, the hashtable is well suitable to support fast data retrieving with a very low computational cost and, ideally it is independent with regard to the number of stored entries.
}

From a configuration point of view, the \textit{inbound} processing just relies on the plain IPv6 routing through \textit{My Local SID Table}: the new route is added with the \texttt{ip -6 route add} command of the \texttt{iproute2} suite, by also specifying the behavior to be activated in the parameters of the command.
\shortver{The details on the configuration commands are in \cite{legacy-vnf-sr-proxy-linux-kernel-extended}.}
\extended{Appendix~\ref{sec:appendix_a} provides further details on the configuration of the \textit{inbound} processing.}

\subsubsection{Auto-learning Process}

The auto-learning process consists in learning the information related of the VNFs chain (i.e., the list of segments in the SRv6 header) from the inbound packets, without the need of a static configuration. The learned information is saved in a \textit{per-netns hashtable}. We have introduced an \texttt{age} parameter to control the rate at which the \textit{per-netns hashtable} can be updated. This parameter can be set during the setup of the LWT routing entry in \textit{My Local SID Table}. When different from 0, the \texttt{age} parameter represents the minimum interval (in seconds) between two write operations in the \textit{per-netns hashtable} for the same VNF. Setting the \texttt{age} to 1 second  corresponds to a maximum reconfiguration delay of 1 second for a NFV node when the VNF chain is changed by an ingress node and this is the default we used in our experiments. If \texttt{age} equals 0, the \textit{per-netns hashtable} is updated for every inbound packets, providing the fastest possible reconfiguration time for a VNF chain. In the performance evaluation section, we have analyzed the performance cost for the continuous updating of the \textit{per-netns hashtable} with respect to the default minimum reconfiguration delay of 1 second.
\extended{The \texttt{age} parameter registers the last time the headers have been updated and it is used also to determine, when a packet is received, if it is the time to replace stale data with new fresh one. The auto-learning operation is performed only during the \textit{inbound} processing. The learned information (VNFs chain) is retrieved during the \textit{fromVNF} processing using the incoming interface\footnote{the current implementation of the \textit{dynamic proxy} assumes that the same interface is used to interact with VNF in the two directions} of the packet to rebuild the whole SRv6 packet ready for being forwarded into the network.
\\
Setting properly the \texttt{age} parameter has an important impact on the performance of the system and a proper trade-off is necessary according to the use case to be supported. In a shared-memory producer-consumer context, we can identify the \textit{inbound} processing as the content producer, and the \textit{fromVNF} one as the consumer. Indeed, the former is in charge of keeping the \textit{per-netns hashtable} up-to-date, while the latter accesses the structure for retrieving the headers. Considering this model, the aging parameter can be seen as the upper-bound of data production/refresh rate. By setting it to the maximum limit, it is possible to prevent overloading of the SRv6 NFV node caused by high-rate writing in the shared memory.
\\
This problem is particularly noticeable in all of those systems based on multi-core architectures: the Linux networking stack allows to assign the received packets to all available computing units in order to process them in parallel and to support high data rates. However, this means that several End.AD processing operations may occur at once and, potentially, they may involve updating the same IPv6 and SRv6 headers. Very frequent and simultaneous shared memory updates by multiple CPUs can lead to conflicts that can negatively affect the overall performance of the system. For all these reasons, small values of the \texttt{age} parameter make the system more responsive to chain (SRv6's segment list) changes, but on the other side they can push heavy and unnecessarily load to the SRv6 NFV node due to high data refresh rate.}

\extended{
\subsubsection{End.AS design}
The End.AD differs from the End.AS just on the way the stored headers are managed. The End.AS behavior is a simplification of the End.AD because it does not need to deal with the auto-learning process. Indeed, it uses chain information which has been saved once during the behavior configuration. The list of segments does not change during the entire life of the End.AS instance unless it is first deleted and then saved with new headers values.
}

\subsubsection{FromVNF Processing}
\label{sec:fromVNF_processing}

The \textit{fromVNF} LWT tunnel is meant to work in tandem with its \textit{inbound} counterpart.
\textit{fromVNF} packets do not carry any SID as it happens for the \textit{inbound} ones. As result, in order to select the correct (\textit{fromVNF}) LWT tunnel and processing each packet accordingly,  we can rely only on the incoming interface between the VNF and the NFV node through which packets come back. Hence, we add an entry in the IPv6 Routing Policy DB (RPDB) for each VNF to be supported. Every RPDB entry is also known as IPv6 rule, as the command used to configure it is \texttt{ip -6 rule}.
The rule points to a different routing table for each VNF, in which there is only a default route, pointing to the LWT tunnel associated to the VNF. This means that for $N$ VNFs, we will have $N$ rules and $N$ routing tables. Figure~\ref{fig:proxy1_fromvnf} provides a representation of the described \textit{fromVNF} processing.
\extended{
Let us analyze with more details the motivation for this design. The \textit{fromVNF} LWT tunnel can not be tied to any route with a specific prefix because the IPv6 packets sent by VNF can use any destination address and do not have any relationship with the SIDs. Moreover, each End.AD \textit{fromVNF} tunnel expects to receive traffic by its own layer-2 interface (\texttt{veth0} in Figure \ref{fig:proxy1}), with no regards about the IPv6 destination address of the packets. This means that, in order to apply the \textit{fromVNF} processing function to an incoming packet, the SRv6 NFV node has to retrieve the route that points to the right LWT tunnel using only the identifier of the interface where such as packet has been received. As a consequence of this, the \textit{fromVNF} End.AD design has to deal with: i) the problem of designating an IPv6 prefix to be used for creating a route pointing to a custom processing function (LWT), and ii) the issue of steering incoming traffic received on a specific interface through such as route.
\\
The first issue can be easily solved by using as route prefix the \textit{any address} which is often indicated by \texttt{``::''}. Generally, the \textit{default route} is selected by the routing algorithm when the IPv6 destination address can not be managed by any other. However, this usage gives rise to a new problem. Indeed, creating a LWT on a \textit{default route} has the side effect that no more than one VNF can be handled by the SRv6 node using a single table. Moreover, control traffic that transits through the SRv6 node and for which there are no given explicit routes may be wrongly handled by the LWT installed on the \textit{default route}. Thankfully, this problem can be easily solved by installing every \textit{default route} into a different IPv6 routing table and creating, for each of these, a rule in the IPv6 Routing Policy DB. 
Such rule is meant to instruct the IPv6 network system to perform route look-up on a specific table based on a specified match. The usage of an IPv6 policy route solves also the issue ii) as at this point we can use the \textit{fromVNF} interface (\texttt{veth0} in the above example) as match and a \texttt{goto-table N} as action predicate. In this way we can relate an interface to a specific \textit{default route} that has attached to a LWT.
\\
Figure~\ref{fig:proxy1_fromvnf} shows an high-level overview of proposed solution with the \textit{fromVNF} LWT tunnel integrated in the IPv6 routing network subsystem. Whenever a plain IPv6 packets, sent by VNF, arrives at SRv6 NFV node, it is handled by the Linux networking stack. After passing the \texttt{pre-routing} stage, the kernel tries to determine the right processing of the packet. It invokes the route look-up operation, but this time the routing algorithm finds first an entry in the RPDB of the node and does not consider IPv6 destination address at first. Indeed, thanks to custom IPv6 rules (one for each \textit{fromVNF} tunnel) the routing algorithm is capable to retrieve the IPv6 table tied to the incoming interface of the packet. At this point, the routing algorithm makes use of this table to find out the route that matches with the received packet. In this specific case, the routing algorithm selects and returns the only route available, the \texttt{default} one, that is attached to a specific End.AD tunnel. Once the plain IPv6 packet has been received by the \textit{fromVNF} processing function, it leverages the identifier of the incoming interface of the packet to search for the popped IPv6 and SRv6 headers within the \textit{per-netns hashtable}. If a result is found, the processing function forges a new packet and sets the headers of such as packet with those that have just been retrieved. The plain IPv6 packet is encapsulated into the newly created one and then the whole packet is delivered towards its destination. This concludes the job of the \textit{fromVNF} LWT tunnel processing operation.}

\subsection{SRNKv2}
\label{sec:proxy2}

After the implementation of SRNKv1, we critically revised its design, by identifying the following main shortcomings: i) two LWT tunnels are used for the two directions \textit{inbound} and \textit{fromVNF} related to the same VNF; ii) a different routing table needs to be instantiated for each VNF so that the correct LWT tunnel can be associated to the \textit{fromVNF} packets; iii) the use of the Linux Policy Routing framework implies to sequentially scan a list of \textit{rules} to identify the VNF interface from which a packet is coming and associate a specific routing table. In particular, the first two shortcoming correspond to a waste of memory resources, while the third one to a waste of processing resources (see Sec.~\ref{sec:sr_perf}).



\begin{figure*}[t]
    \centering
    \includegraphics[width=16.5cm, height=6.5cm]{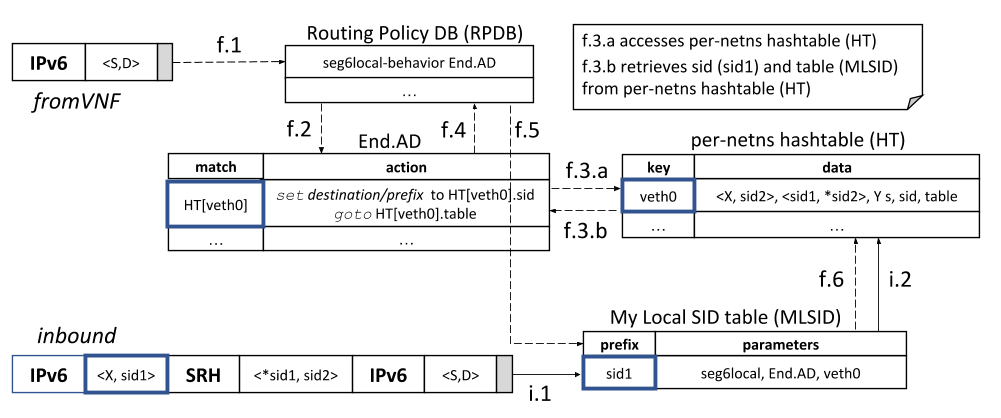}
    \caption{SRNKv2 design}
    \label{fig:proxy2}
    \vspace{-3ex}
\end{figure*}


The revised design of SRNKv2 is shown in Figure~\ref{fig:proxy2}. The most important improvement is an extension to the Linux Policy Routing framework, in order to avoid the linear scan of the list of rules to match the VNF interface (i.e. one rule for each VNF). A new type of IPv6 rule, called SRv6 extended rule, is added and used in the \textit{fromVNF} processing (f.1). The new rule (indicated as seg6local-behavior End.AD in the Figure~\ref{fig:proxy2}) is added to the Routing Policy DB. The \textit{selector} of this rule performs the lookup of the packet incoming interface (f.2) into the \textit{per-netns hashtable} that includes all the VNF interfaces handled by the \textit{dynamic} proxy. In this way, it is possible to understand if the packet is coming from a VNF and to retrieve the needed information (f.3.a, f.3.b). The SID associated to the VNF is used to perform the search (f.5) into the \textit{My Local SID Table}, which returns the associated tunnel. 

In the new design, there is actually a single LWT tunnel for the two directions. In fact, the code that is executed when a routing entry points to the tunnel is able to understand if the packet belongs to the \textit{inbound} or to the \textit{fromVNF} direction and behave accordingly. Thanks to the lookup in the \textit{per-netns hashtable}, which allows to retrieve the SID associated with the VNF, it is not needed anymore to have a separate routing table for each VNF.

\extended{
\subsection{Implementation of other SR proxy types} 
In addition to the implementation of the SR \textit{dynamic proxy}, we have already implemented the static proxy behavior. This is actually a simple extension of the dynamic one (we only needed to develop command line static configuration tools and disable the ``learning'' of segment list from \textit{inbound} packets. In principle, our SR proxy solution can be extended to implement also the masquerading proxy, but this is currently for further work.}

%% file: inc/04-test-environm.tex
\section{Testing Environment}
\label{sec:sr_test}
\subsection{Testbed Description}
\begin{figure}
    \centering
    \includegraphics[width=0.44\textwidth]{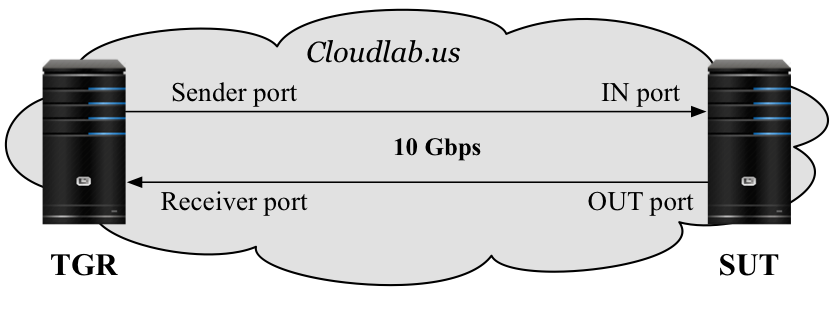}
    \caption{Testbed scenario}
    \label{fig:testbed}
    \vspace{-3ex}
\end{figure}

We built our testbed according to RFC 2544 \cite{rfc2544}, which provides the guidelines for benchmarking network interconnecting devices. Figure~\ref{fig:testbed} shows the considered scenario. More in depth, the testbed is composed of two nodes, which we call respectively \textit{Traffic Generator and Receiver (TGR)} and \textit{System Under Test (SUT)}. Both \textit{TGR} and \textit{SUT} have two ports. In our experiment we consider the traffic crossing the SUT in one direction. As a consequence, the ports can be identified as in Figure~\ref{fig:testbed}: the traffic is generated by the TGR on its Sender port, enters the SUT from the IN port, exits the SUT from the OUT port and then it is received back by the TGR on its Receiver port. In this configuration, the \textit{TGR} can easily perform all different kinds of statistics on the transmitted traffic including packet loss, delay, etc. 

The testbed is deployed on CloudLab \cite{ricci2014introducing}, a flexible infrastructure dedicated to scientific research on the future of Cloud Computing. In our experiment, each node is a bare metal server with Intel Xeon E5-2630 v3 processor with 16 cores (hyper-threaded) clocked at 2.40GHz, 128 GB of RAM and two Intel 82599ES 10-Gigabit network interface cards. The SUT node hosts the SR-unaware VNF, which is implemented as a Linux network namespace.

The SUT machine is running a compiled version of Linux kernel 4.14 patched with our End.AD proxy behavior implementations (namely SRNKv1 and SRNKv2). It has also a modified version of iproute2 tool \cite{iproute2}, which allows the configuration of the \textit{dynamic} proxy. Focusing on the traffic generator, we exploit TRex \cite{trex-cisco} in the TGR node. TRex is an open source traffic generator powered by DPDK \cite{dpdk}. We have used the described testbed scenario also in \cite{ahmedperformance}, which provides further details on the nodes configuration for correct execution of the experiments.
\subsection{Methodology}

\label{sec:methodology}
From the TGR node, we generate SRv6 traffic using TRex. We consider IPv6 UDP packets encapsulated in outer IPv6 packets. The outer packets have an SRH with a SID list of two segments. The first SID points to the SR-unaware VNF running in the SUT, the second SID corresponds to the Receiver interface of the TGR node from the point of view of the SUT. Regarding the packet size, we have followed the indications from the Fast Data I/O Project (FD.io) Continuous System Integration and Testing (CSIT) project report \cite{csit-report}. In particular, the inner IPv6 packets have an UDP payload of 12 bytes, corresponding to 60 bytes at IPv6 level. The SR encapsulation adds 40 bytes for outer IPv6 header and 40 bytes for the SRH with two SIDs. The Ethernet layer introduces 18 bytes for Ethernet header and CRC, plus 20 bytes at the physical layer (preamble, inter frame gap). TRex takes in input a file with the dump of a sample packet and can reply the packet with a configurable packet rate, denoted as $PS$ rate [kpps] (Packet Sending rate). Interestingly, $PS$ can be configured with relatively high precision.

\begin{figure}
    \centering
    \includegraphics[width=0.485\textwidth]{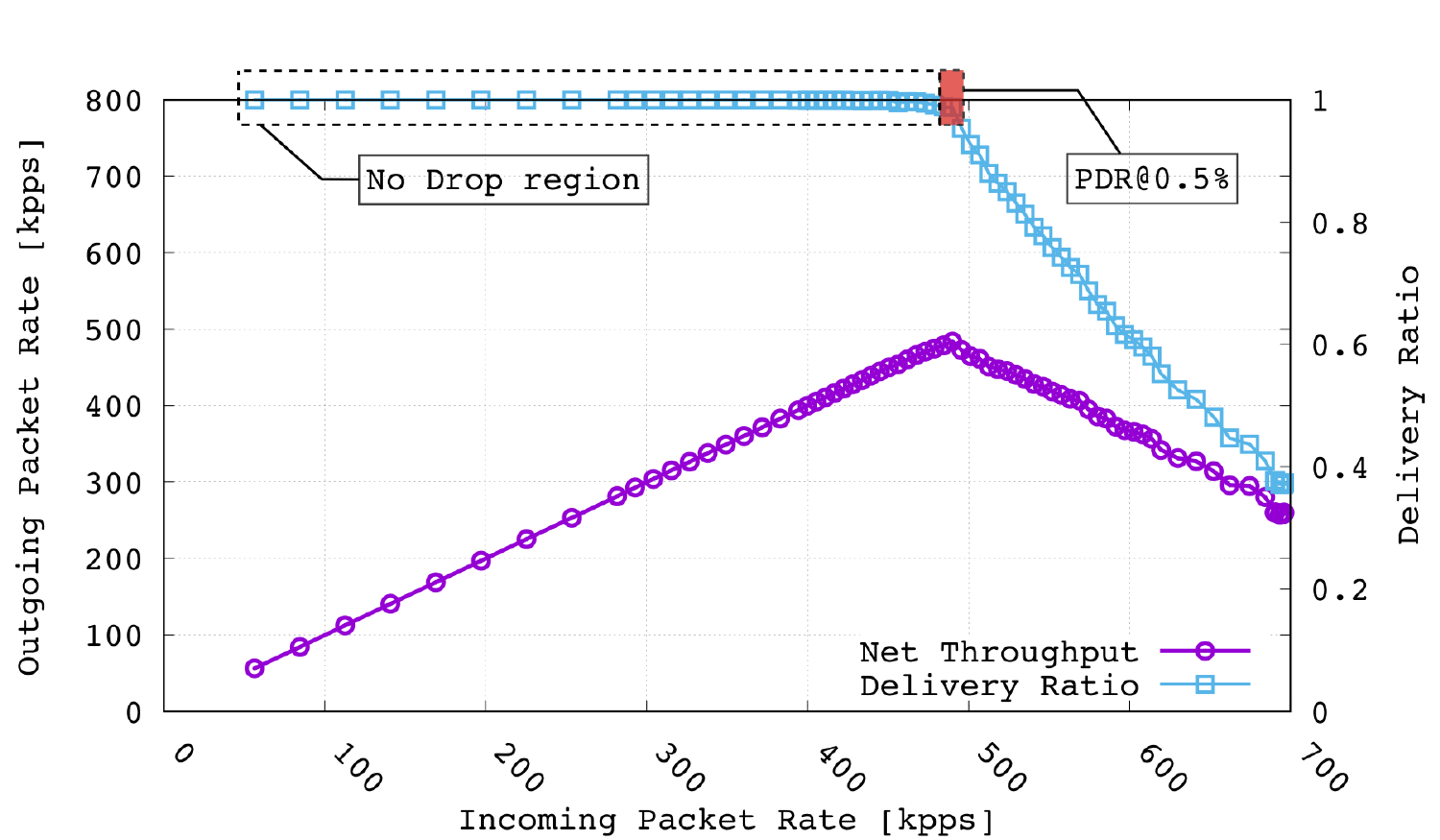}
    \caption{Throughput, Delivery Ratio, PDR}
    \label{fig:no-drop-region}
    \vspace{-4ex}
\end{figure}

For a given configuration of the SUT, we characterize the SUT performance by generating traffic at a given $PS$ rate [kpps] for a duration $D$ [s] (usually $D=10 s$ in our experiments). Let the number of packets generated by the TGR node and incoming to the SUT in an interval of duration $D$ be $P_{IN}$ (Packets INcoming in the SUT). We define the number of packets transmitted by the SUT (and received by the TGR) as $P_{OUT}$ (Packets OUTgoing from the SUT). The throughput $T$ is $P_{OUT}/D$ [kpps]. We define the Delivery Ratio (DR) as $P_{OUT}/P_{IN}=P_{OUT}/(PS*D)=T/PS$. We run a number of test repetition (e.g. 15) to evaluate the average and standard deviation, then we replicate the measurements for different sending rates. We are assuming that the performance is limited by the processing capacity of the SUT, in our experiments we make sure that a single CPU is used for the forwarding of the packets. In particular the same CPU is used for the operation of the NFV node and of the VNF. An example of result is show in Figure~\ref{fig:no-drop-region}, for the \textit{baseline} case that will be described in the next section. For each $PS$ rate reported in the X axis, we plot the Throughput [kpps] (right Y axis) and the Delivery Ratio (left Y axis) as averages of the 15 repetitions. The standard deviation is not shown in the figure, because it is so close to the average that cannot be distinguished. Starting from the left (low $PS$) there is a region in which the Throughput increases linearly with the $PS$ and the Delivery Ratio is 1. Ideally, the Delivery Ratio should remain 1 (i.e. no packet loss) until the SUT saturates its resources and starts dropping a fraction of the incoming packet. This is defined as the \textit{No Drop} region and the highest incoming rate for which the Delivery Ratio is 1 is defined as \textit{No Drop Rate} (NDR). On the other hand, in our experiments with the Linux based SUT we measured very small but not negligible packet loss ratio in a region where we have an (almost) linear increase of the Throughput. Therefore, according to \cite{csit-report} we define a Partial Drop Rate (PDR) by setting a threshold for the measured Loss Ratio, typically we used 0.5\% as threshold corresponding to a Delivery Ratio of 0.995. The PDR@0.5\% is the highest rate at which the Delivery Ratio is at least 0.995. The usefulness of the PDR is that it allows to characterize a given configuration of the SUT with a single scalar value, instead of considering the full relation between Throughput and Incoming rate shown in Figure~\ref{fig:no-drop-region}. When the incoming rate overcomes the PDR, the throughput starts to decrease due to trashing phenomena. 
Actually, finding the PDR of a SUT configuration is in general a time consuming operation as it is needed to scan a broad range of possible traffic rates. For this reason, we designed and developed a PDR search algorithm that optimizes the time needed to find the PDR with a given precision (details in~\cite{ahmedperformance}).

%% file: inc/05-performance-analysis.tex
\begin{figure}[h!tb]
    \centering
    \includegraphics[width=0.48\textwidth]{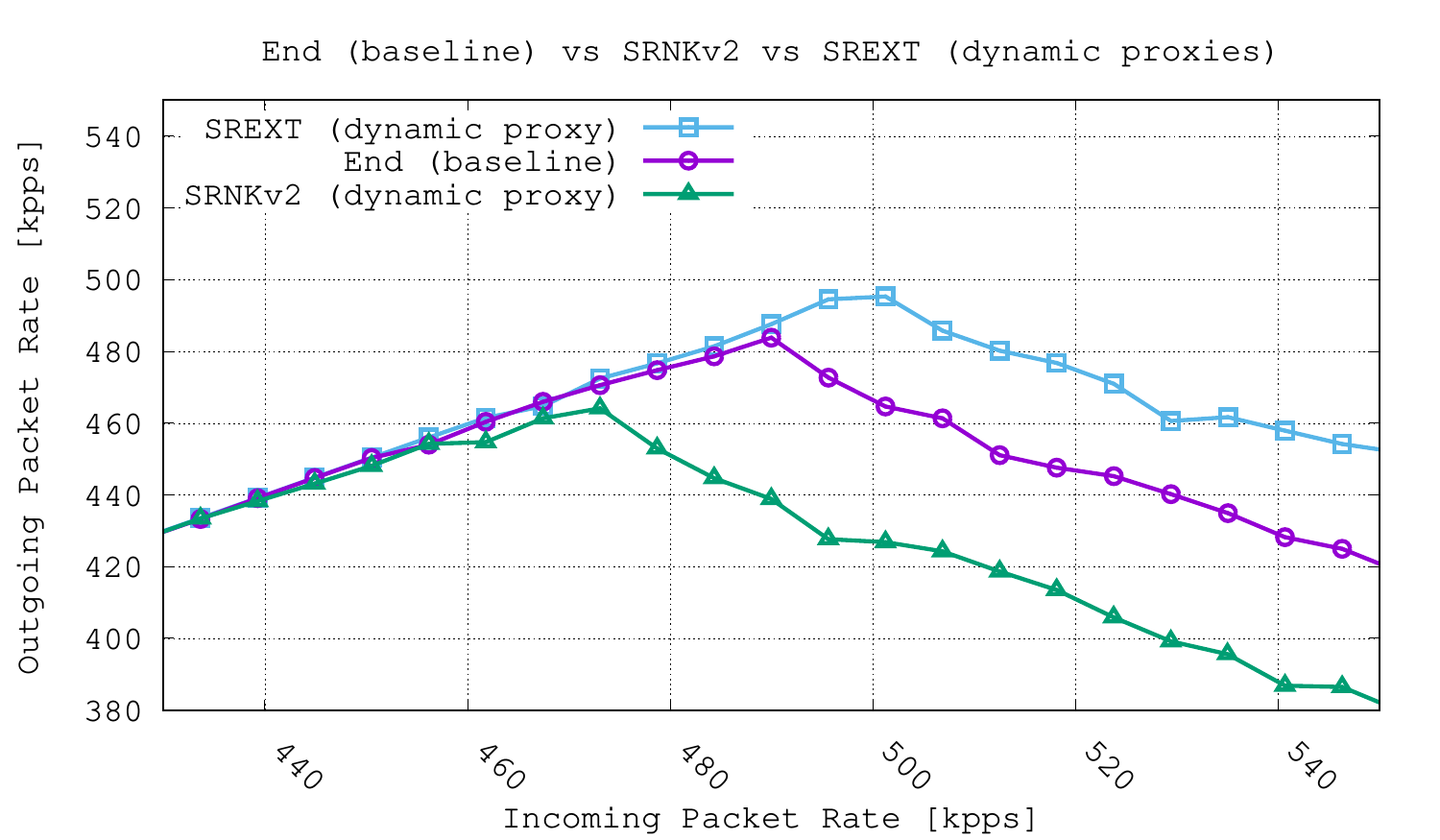}
    \caption{Comparison of End (baseline) vs SRNKv2 vs SREXT}
    \label{fig:testbed_4}
    \vspace{-1ex}
\end{figure}

\section{Performance Analysis}
\label{sec:sr_perf}

Figure~\ref{fig:testbed_4} reports the performance characterization of our solution (SRNKv2) compared with a baseline reference (End) and with the pre-existing solution (SREXT). The traffic pattern used for the characterization has been described in subsection~\ref{sec:methodology}. In the baseline scenario, no Segment Routing behavior is configured in the NFV node that simply forwards the \textit{inbound} and \textit{fromVNF} packets according to IPv6 destination addresses. On the other hand, the VNF is SRv6 aware and performs the so-called SRv6 \textit{End} behavior (for this reason the scenario is called End). In the End behavior, a node receives the packets destined for itself and advances the pointer of the segment list to the next segment, changing the IPv6 destination address. As a result, in the baseline scenario the SUT performs two regular IP forwarding operations (each one with a lookup in the routing table) and one SRv6 End behavior (which include a lookup for the incoming SID, an SRv6 header processing and a lookup for the next segment). In the SRNKv2 case, the SUT performs a routing lookup for the incoming SID, it decapsulates the packet according to the \textit{dynamic} proxy behavior and forwards it to the VNF. The VNF performs a plain forwarding operation (routing lookup) on the inner packet. Moreover, the match on the incoming interface is performed in the NFV node when receiving the packet. The packets are re-encapsulated after retrieving the proper header and finally an IPv6 forwarding operation is performed. The SREXT operations are similar to the ones in the SRNKv2 scenario. The main difference is that the matching on the \textit{inbound} packets is not performed in the Linux IPv6 forwarding/routing but the packets are captured in the pre-routing phase. Therefore, the regular forwarding operations are skipped, leading to an higher performance. The PDR@0.5\% for SRNKv2, baseline and SREXT are reported in Table~\ref{table:throughput_comparison}. Our SRNKv2 implementation, which also perform decapsulation and re-encapsulation of packets, shows only a 3.7\% performance degradation with respect to the baseline forwarding. The SREXT module, which skips the Linux routing operations by capturing packets in the pre-routing hook has a forwarding performance boost of 2.4\% with respect to baseline forwarding.

The simplification in routing operations introduced by the SREXT module should be taken into account when making a performance comparison with the in-kernel \textit{dynamic} proxy variants.
The fact that an external module outperforms the in-kernel solution is not surprising in this case. The SREXT logic is tailored to specifically handle the SRv6 case one. Therefore, it can cut off all the generic code that is normally needed to determine the fate of each single packet as well as the protocol handler that should be called to process the data. Indeed, both versions of SRNK have to waste CPU cycles on possible \textit{netfilter} hooks and rule lookup before being able to handle a SRv6 packet during the routing operation in Linux kernel. So, performance penalty of SRNK is not  the result of a poor design. Instead, it is the side effect of the design choice of the Linux kernel networking stack which wants to be as generic as possible for dealing with a very broad range of different protocols. 

\begin{table}[ht]
\caption{Throughput (PDR@0.5\%) in kpps}
\label{table:throughput_comparison}
\centering 
\begin{tabular}{ |m{2.0cm} |m{2.0cm} |m{2.0cm} |} 
\hline
SRNKv2 & Baseline (End) & SREXT\\ 
\hline
444.2 & 461.1 & 472.3 \\ 
\hline
\end{tabular}
\label{table:pdr-throughput-comp} 
\end{table}

\begin{figure}[h!tb]
    \centering
    \includegraphics[width=0.48\textwidth]{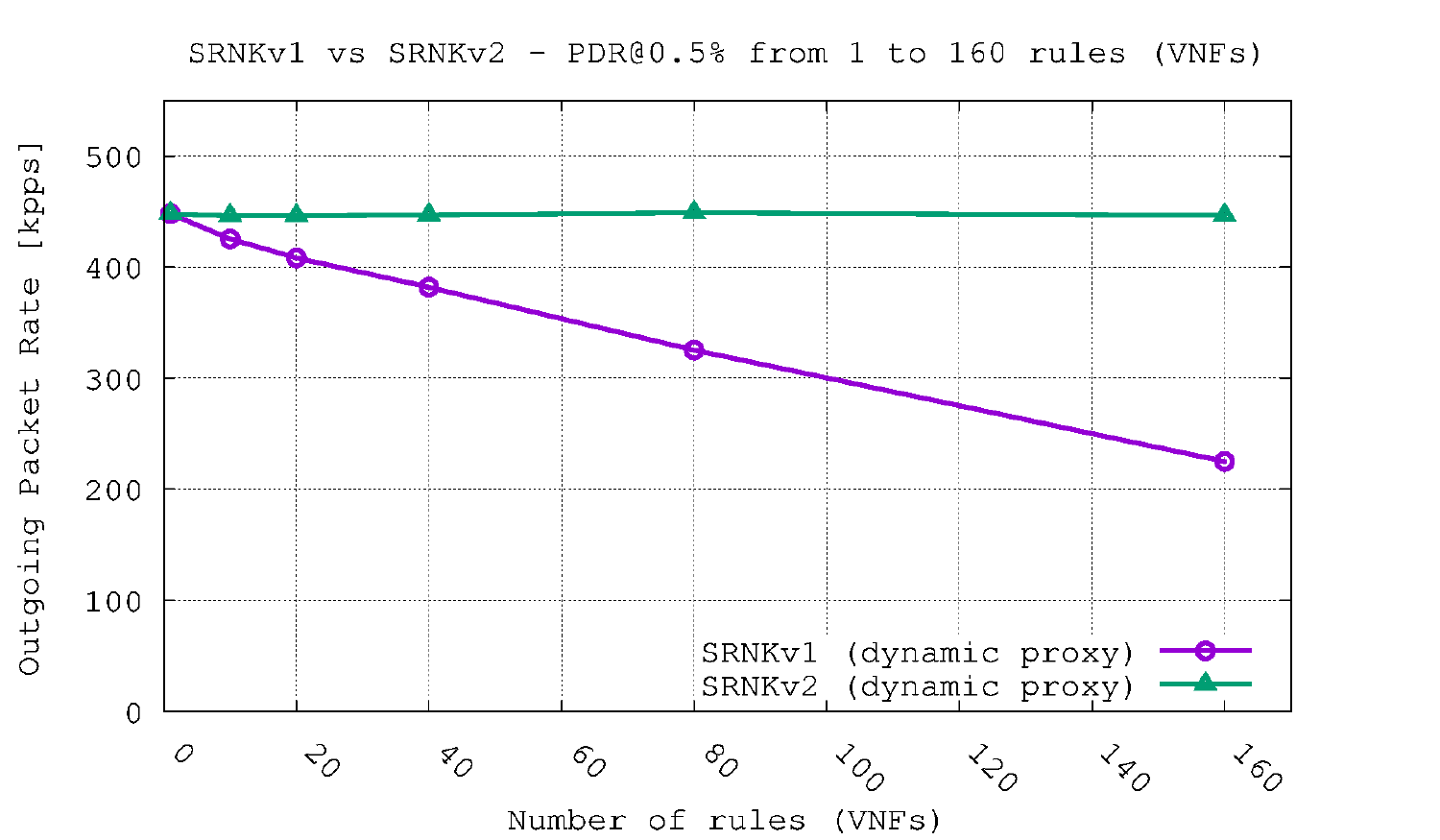}
    \caption{PDR@0.5\% vs. number of rules}
    \label{fig:testbed_num_rules}
    \vspace{-3ex}
\end{figure}

Figure~\ref{fig:testbed_num_rules} analyzes the poor scalability of our first design (SRNKv1) based on the regular Linux Policy Routing framework. We show the PDR@0.5\% versus the number of Policy Routing rules that are processed before the matching one. Consider that the number of rules corresponds to the number of VNFs to be supported and that the rules are scanned sequentially until the matching one. Therefore, the performance with $N$ rules can be read as the worst case performance when $N$ VNFs are supported, or as the average case with $2N$ VNFs (because $N$ rules needs to be checked on average with $2N$ VNFs). A linear degradation of the performance with the number of rules is clearly visible, for example when there are 80 rules the PDR@0.5\% is 28.4\% lower than the PDR@0.5\% for SRNKv2 or for SRNKv1 with a single rule (for 160 rules the PDR@0.5\% is 50.6\% lower).

As for the impact of the auto learning feature of the \textit{dynamic} proxy, in all the experiments shown so far we have evaluated the SRNK performance by setting the \textit{age} parameter to 1 second (hence limiting the update rate to one update/s). We run an experiment by setting it to 0 (no limitation on the update rate, so that the VNF chain is updated for each incoming packet). Under this condition, we were not able to consistently achieve a delivery ratio higher than 0.99 even for low packet rates. Therefore, we evaluated the PDR@2\%, which was around 392 kpps, if we compare to the PDR@0.5\% (444.2 kpps) of SRNkv2 with aging 1 second, we can estimate a decrease of performance not less than 11\% for updating the VNF chain at the highest possible rate (i.e. for every incoming packet). 

Finally, we analyzed the cost of performing the interface lookup with the new extended SRv6 policy rule, separately from the decapsulation and encapsulation operations which are executed in the SRNKv2 scenario. There are two motivations for this analysis. First the policy rule needs to be executed for all IPv6 packets, so it introduces a performance degradation also for non-matching packets that it worth to be evaluated. The second reason is that the proposed mechanism could be reused in scenarios with multiple policy rules based on the incoming interfaces. This would require an extension to the Linux Policy Routing framework, the performance evaluation is a part of the cost-benefit analysis for this extension. For this performance analysis, we start from the  baseline (End) scenario in which the packets are only forwarded in the NFV according to plain IPv6. We consider two scenarios: i) \textit{Ext. SRv6 Rule} with an extended SRv6 rule with no matching interfaces, the rule will be checked for all \textit{inbound} and \textit{fromVNF} packets; ii) \textit{80 Plain Rules} with 80 rules which try to match an interface are added (with no matching interfaces), these 80 rules will be checked for all \textit{inbound} and \textit{fromVNF} packets. The PDR@0.5\% is reported for the baseline and the described scenarios. The performance degradation in the packet forwarding for adding the lookup with the extended SRv6 rule is only 2.7\%. On the other hand, adding 80 policy rules implies a big degradation of the forwarding performance, which becomes 28.1\% smaller.

\begin{table}[ht]
\caption{Extended SRv6 policy rule performance}
\label{table:Policy_routing_new}
\centering 
\begin{tabular}{ |m{2.35cm} |m{1.25cm} |m{1.8cm} |m{1.66cm} |} 
\hline
 & Baseline & Ext. SRv6 Rule & 80 Plain Rules\\ 
\hline
PDR@0.5\% [kpps] & 461.1 & 448.5 & 328.0 \\ 
\hline
Performance penalty & - & 2.7\% & 28.1\% \\ 
\hline
\end{tabular}
\label{table:policy-routing} 
\end{table}
\vspace{-3ex}

%% file: inc/06-related-work.tex
\section{Related work}
\label{sec:rel_work}

\subsection{Service Function Chaining}
Network Service Header \cite{nsh-id} (NSH) has been proposed as solution to implement the SFC architecture defined in \cite{RFC7665}, which specifies that the \textit{SFC encapsulation} transports all the necessary information to map the packets to a specific sequence of \textit{Service Functions} (VNFs) that will process the packets along the path. Segment Routing Header (SRH) is inline with the SFC architecture defined in \cite{RFC7665}. Moreover, it offers optional TLVs (Type-Length-Value) to carry on additional information (like NSH metadata). The most important difference with respect to the SRv6 solution is the need of state information in the SFF forwarders  \shortver{(see more in \cite{legacy-vnf-sr-proxy-linux-kernel-extended}).} \extended{(further discussion on the NSH solution and its differences with SRv6 can be found in Appendix~\ref{sec:ext_nsh_tutorial}).} 

\subsection{SRv6 implementations}
From kernel 4.10, Linux supports SRv6 processing including also the implementation of several local behaviors. However, at the time of writing there is lack of support of proxy behaviors. As already mentioned in Section \ref{sec:srext}, the SREXT module \cite{srext-srv6-net-prog} provides a complementary implementation of SRv6 in Linux based nodes. A specific shortcoming of the SREXT module is that it does not support the Linux network namespaces. Therefore, it cannot coexist with the frameworks and tools that rely on network namespaces (e.g. Linux Containers, Dockers...). More in general, an external kernel module is not able to directly access most of the internal structures and usually it is necessary to re-implement them with the risk of realizing inefficient implementations and risking to introduce bugs in the code. The goal of our SRNK implementation is to be integrated in the Linux kernel mainline, so that it can evolve and be maintained together with the Linux kernel.

Another SRv6 implementation is included in the VPP (Vector Packet Processing) platform, which is the open source version of the Cisco VPP technology. The open source VPP is developed under the umbrella of the FD.io project \cite{fdio}. VPP implementation of SRv6 supports most of the behavior defined in \cite{srv6-net-prog} including also the \textit{dynamic} proxy behavior. As reported in \cite{csit-report}, the forwarding performance of VPP is in general very high. For example a NDR (No Drop Rate) of around 5.5 Mpps is reported for a \textit{dynamic} proxy setup similar to SRNK. These VPP performance evaluations are not directly comparable with our measurements, because they only focus on the SR-Proxy operations, while we have included the processing inside the VNF in our evaluation. In any case with comparable setups, VPP will outperform our implementation (as VPP also outperforms Linux kernel forwarding). Nevertheless, there is still value in enhancing the SRv6 functionality in the Linux kernel as we have proposed, because VPP is not ubiquitously deployed and there are scenarios in which it is simpler to use a kernel based feature rather than depending on an external framework (more discussion in \cite{legacy-vnf-sr-proxy-linux-kernel-extended}). \extended{VPP is built on top of DPDK \cite{dpdk}. Systems, willing to leverage such framework, need to use DPDK compatible NICs. Furthermore, a very broad number of embedded devices, used as network nodes, do not have the capability to run such frameworks (for example resource constrained devices) or run Linux distributions (for instance LEDE \cite{lede} or OpenWrt \cite{openwrt}) without the support of kernel bypass functionality. Last but not least, contrary to the Linux kernel, DPDK and thus VPP require reserved memory (hugepages for DPDK) and CPUs that could not be used for any other tasks.}

%% file: inc/07-conclusions.tex
\section{Conclusions}
\label{sec:conclusions}

In this paper, we have described SRNK, a \textit{dynamic} proxy for Linux that supports Service Function Chaining based on IPv6 Segment Routing for legacy VNFs, a use case of great importance for service providers. The SRNK implementation is open source (available at \cite{srnk-home-page}) and extends the current Linux kernel implementation of the SRv6 network programming model. SRNK is well integrated in Linux ecosystem, as it can be configured through the well known iproute2 \cite{iproute2} utility. We plan to submit the SRNK code to the Linux kernel mainline. 

We have thoroughly analyzed several performance aspects related to our implementation of the \textit{dynamic} SRv6 proxy. We went through two design and implementation cycles, referred to as SRNKv1 and SRNKv2. We identify a scalability issue in the first design SRNKv1, which has a linear degradation of the performance with the number of VNFs to be supported. The root cause of the problem is the Linux Policy Routing framework. The final design SRNKv2 solved the problem, by introducing an new type of rule in the Policy Routing framework. This rule, called extended SRv6 rule, allows using an hash table to associate an incoming packet with its interface, verifying if the packets is coming from a legacy VNF and retrieving the information needed by the \textit{dynamic} SRv6 proxy to process the packet (e.g. re-encapsulating it with the outer IPv6 header and the Segment Routing Header).

\extended{
As a final remark, we want to stress that even if the SRNKv2 performs a little less than the SREXT dynamic proxy implementation (< 6\%), it is well integrated in the kernel code and it could be maintained with less effort with respect to the external module. SREXT takes its advantage on SRNK from the fact that it is highly designed to cut off most of the generic code that slows down the kernel networking performance by hooking itself directly in the prerouting path. Anyway, SREXT is not capable to deal with network namespaces, it can not access to some crucial internal structures and unexported functions (such as the ones offered by the kernel used for encap/decap SRv6 packets) and it does not rely on a standard configuration tool in userspace (whereby SRNK is well integrated with a patched version of iproute2). SRNK should be thought as an effort to bring inside the Linux kernel an extra feature for supporting legacy applications and leveraging the power of network programming introduced by SRv6.}




%% file: inc/appendix_a.tex
\appendices

\section{SRNKv1 configuration}
\label{sec:appendix_a}

Hereafter we report the configuration procedure of the SRNKv1 implementation.

\subsection{Inbound processing}

The \textit{inbound} End.AD tunnel leverages only the classic IPv6 routing. With reference to the testbed network which has been depicted in Figure \ref{fig:testbed}, each packet with destination \texttt{fdf1::2} that comes into SFF/SR node from any interface will be handled by the LWT installed on that route. In order to create an \textit{inbound} End.AD tunnel, which is able to handle packets destined to the VNF with SID \texttt{fdf1::2}, the command used is:

\begin{Verbatim}[frame=single,fontsize=\small]
$ ip -6 route add fdf1::2/128 \
encap seg6local action End.AD chain inbound \
oif veth0 nh6 fdf1::2 age 5 \
dev veth0
\end{Verbatim}

It is easy to identify in its structure three different parts:

\begin{itemize}
\item \texttt{ip -6 route add fdf1::2/128} is used to add the route \texttt{fdf1::2} in the IPv6 routing tables;
\item \texttt{encap seg6local} is used for specifying to the IPv6 networking subsystem to create a \texttt{seg6local} tunnel which handles packets for \texttt{fdf1::2}
\item \texttt{action End.AD chain inbound oif veth0 nh6 fdf1::2  age 5} is used for specifying the behavior of the \texttt{seg6local} tunnel.
\end{itemize}

In this sub-command, the action is defined as End.AD and the direction of the data flow is towards the VNF (\texttt{chain inbound}). As we explained in Section \ref{fig:proxy1_inbound}, each packet that comes into this tunnel is subjected to an outer IPv6 and SRv6 headers de-capsulation. The \texttt{nh6} param is used to inform the \textit{inbound} tunnel about the next hop to which each packet has to be sent, and in this case is the SID of the legacy VNF.

\subsection{FromVNF processing}

The \textit{fromVNF} End.AD tunnel has to perform the inverse operation realized by the \textit{inbound} counterpart. It has to restore the IPv6 and SRv6 headers using only the incoming interface of the packet as key search. At this point, we need to instruct the network subsystem on how it should send traffic to the right \textit{fromVNF} tunnel. Using the \texttt{ip -6 rule} tool we are able to manipulate the Routing Policy DB of the nodes and issue a command that informs the system to make use of a given IPv6 routing table when traffic arrives at some specific interface. The \texttt{ip rule} command is the following:

\begin{Verbatim}[frame=single]
$ ip -6 rule add iif veth0 table 100
\end{Verbatim}

After the execution of this command, every time a packet arrives at the ingress interface \texttt{veth0}, the routing system will try to find a route in \texttt{table 100} that matches with the destination address of that packet. 

Instead, to create an \textit{fromVNF} End.AD tunnel on router SFF/SR with the purpose of managing packets coming from the legacy VNF at interface \texttt{veth0}, the following command is used:

\begin{Verbatim}[frame=single]
$ ip -6 route add default \
    encap seg6local action End.AD \
    chain fromVNF iif veth0 \
    dev veth0 table 100
\end{Verbatim}

Also for the case of \textit{inbound} tunnel creation, the ip command can be seen split into three different parts:

\begin{itemize}
\item \texttt{ip -6 route add default} is used to add the \texttt{default route} \texttt{``::''} in the IPv6 routing table \texttt{100};
\item \texttt{encap seg6local} is used for specifying to the IPv6 network subsystem to create a \texttt{seg6local} tunnel which handles packets destined for \texttt{default} address;
\item \texttt{action End.AD chain fromVNF iif veth0} is used to specify the attributes of the behavior that is intended to be created. 
\end{itemize}

The action is End.AD and the direction of the data flow is specified by the chain attribute which is set to \texttt{chain fromVNF}. The \texttt{iif} keyword indicates packets coming from interface \texttt{veth0}. This means also that the tunnel is allowed to listen for incoming traffic only from the interface specified by \texttt{iif}. If it receives packets from another interface, packet is discarded automatically.

%% file: inc/appendix_b.tex
\section{SRNKv2 configuration}
\label{sec:appendix_b}

Hereafter we report the configuration procedure of the SRNKv2 implementation.

\subsection{Tunnel creation}
In our second design (Section \ref{sec:proxy2}) we have introduced the notion of bi-directional tunnel within the definition of the End.AD behavior. The creation of the desired behaviour can be achieved through the following command:

\begin{Verbatim}[frame=single]
$ip -6 route add fdf1::2/128 \
    encap seg6local action End.AD \
    oif veth0 nh6 fdf1::2 age 5 \
    dev veth0
\end{Verbatim}

As it is possible to appreciate, the command closely resembles the one that we have used in the Section \ref{sec:appendix_a} to setup the inbound tunnel, but this time, we are creating only one tunnel that manages the inbound/fromVNF processing for a given VNF.

With reference to the scenario depicted in Figure \ref{fig:testbed}, traffic that arrives at SFF/SR node with destination the VNF's SID (\texttt{fdf1::2}) on any interface except \texttt{veth0} is sent to seg6local LWT associated with the End.AD behaviour for de-capsulation purposes (inbound processing). After that, packets are delivered to VNF using the output interface (for short called \texttt{oif}) \texttt{veth0}. On the other side, traffic that arrives from the interface \texttt{veth0} has to be redirected to the right End.AD tunnel for the encapsulation (fromVNF processing). To accomplish that, the IPv6 rule, described in the next paragraph, is used.

\subsection{IP rule configuration}

The idea behind RPDB configuration relies on the ability to forward packets from a specific incoming interface to the associated End.AD LWT and process them similarly to the inbound process. With this idea in mind we have designed, and implemented the changes described in Section \ref{sec:proxy2} and the following \texttt{ip rule} command to redirect traffic from VNFs towards the right End.AD tunnels for fromVNF processing:

\begin{Verbatim}[frame=single,fontsize=\small]
$ ip -6 rule add seg6local-behaviour End.AD
\end{Verbatim}

The command allows us to add a rule to the RPDB of the node, \texttt{seg6local-behaviour End.AD} is used to specify the local SRv6 behaviour that should be taken into account when the rule is picked up. If the priority of the above rule is not superseded by the priority of other rules, the extended RPDB is able to compare the packet's incoming interface with the \texttt{oif} of the existing LWTs through the \textit{per-netns hashtable}. This means that: if the inbound interface is equals to (exactly) one \texttt{oif},  the patched IP rule subsystem gets the correspondent VNF's SID that is used as the destination address for the packet (in place of the real's one) during the IPv6 routing lookup on ``My Local SID Table''. As soon as the route is resolved, the associated End.AD tunnel is also retrieved and it is exploited for applying fromVNF operations on the incoming packet. Otherwise, if the packet's incoming interface does not belong to any End.AD instance, the packet is treated as usual and the route lookup is performed, by default, on the destination address.

%% file: inc/extended-results.tex
\section{Detailed experiments results}
\label{sec:ext_exper_res}

\begin{figure}[h!tb]
    \centering
    \includegraphics[width=0.48\textwidth]{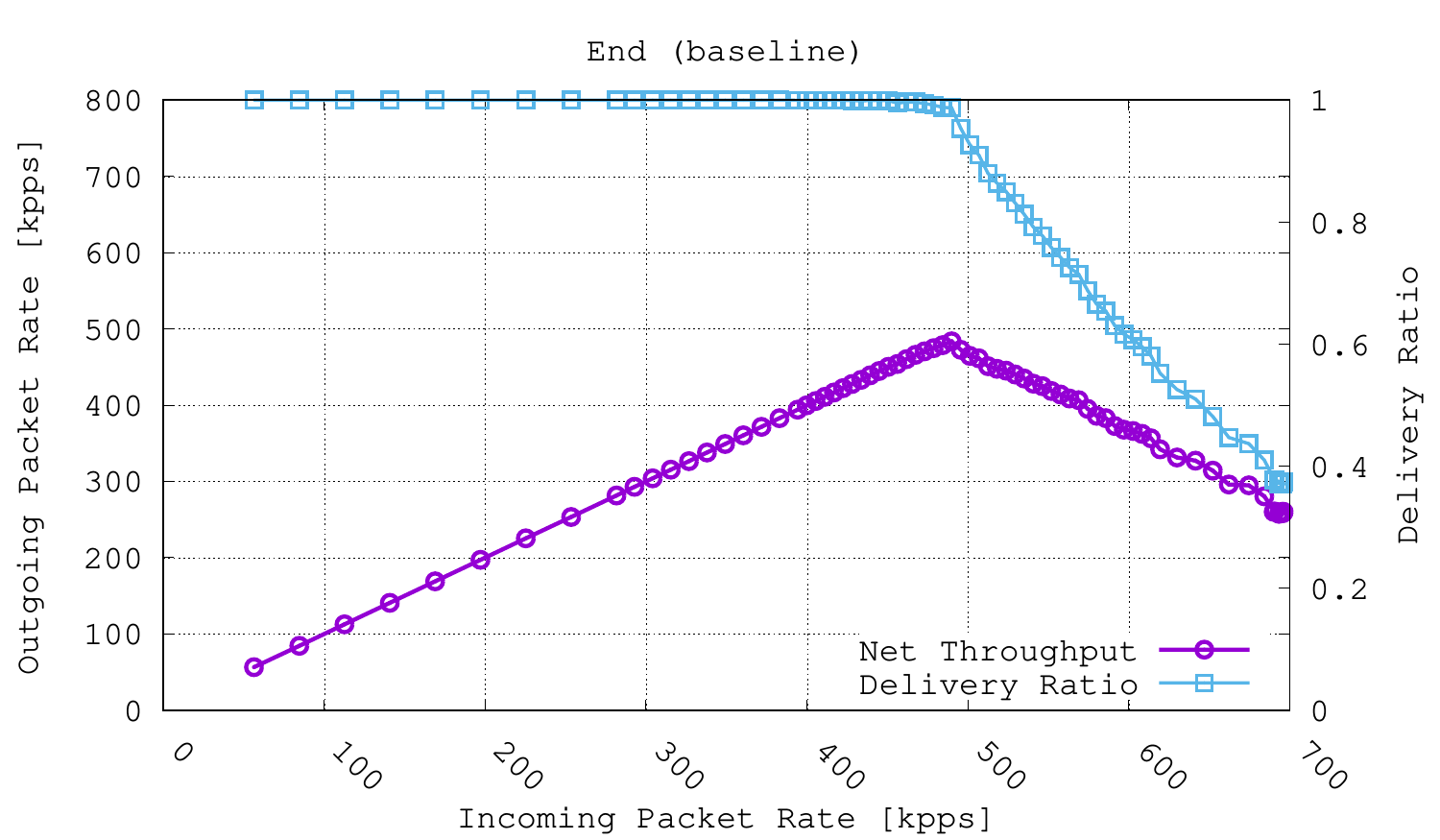}
    \caption{Throughput vs Delivery Ratio for the SR End behaviour (baseline)}
    \label{fig:testbed_1}
    \vspace{-3ex}
\end{figure}

\begin{figure}[h!tb]
    \centering
    \includegraphics[width=0.48\textwidth]{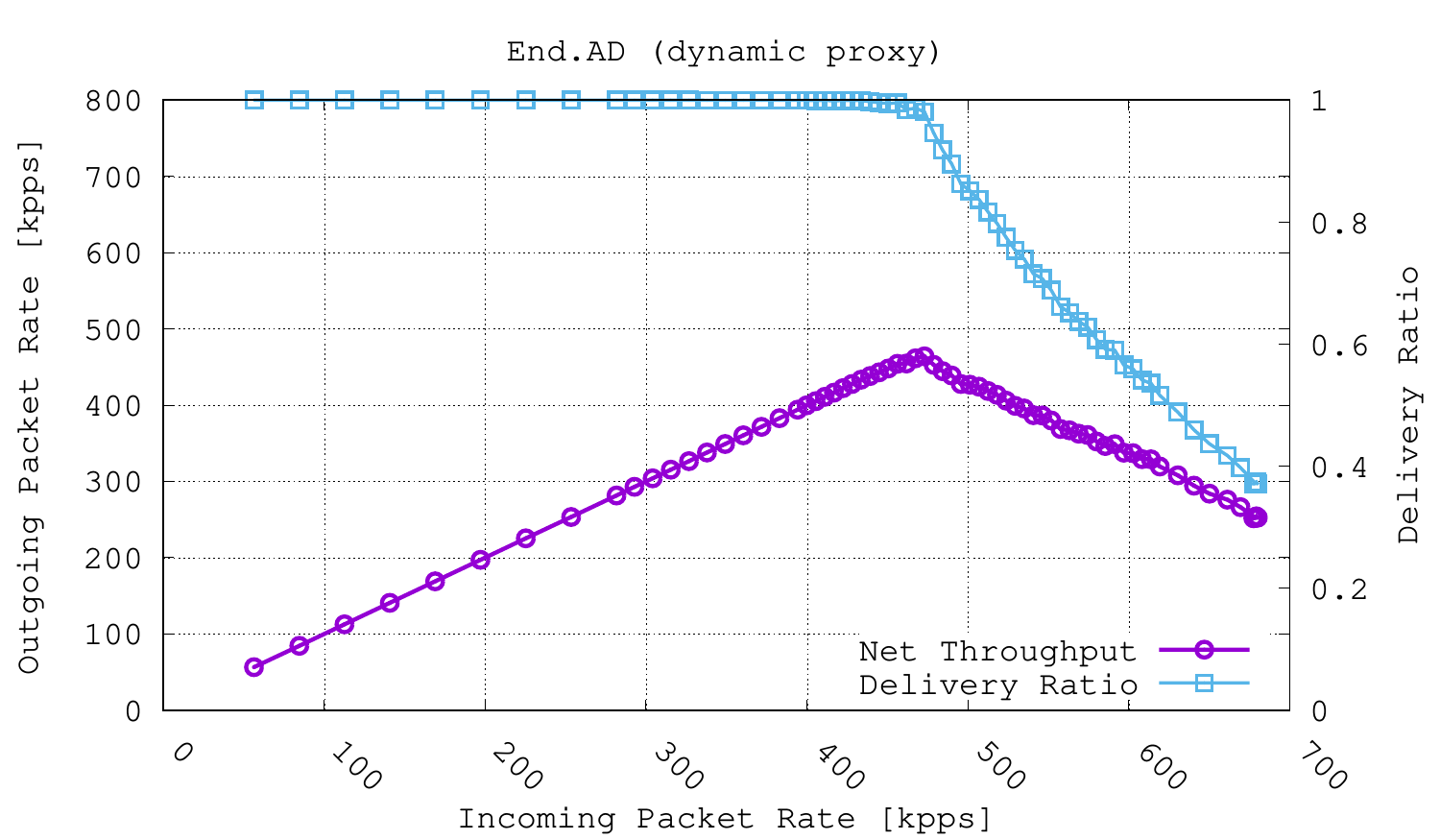}
    \caption{Throughput vs Delivery Ratio for the SR End.AD behaviour (SRNKv2)}
    \label{fig:testbed_2}
    \vspace{-3ex}
\end{figure}

\begin{figure}[h!tb]
    \centering
    \includegraphics[width=0.48\textwidth]{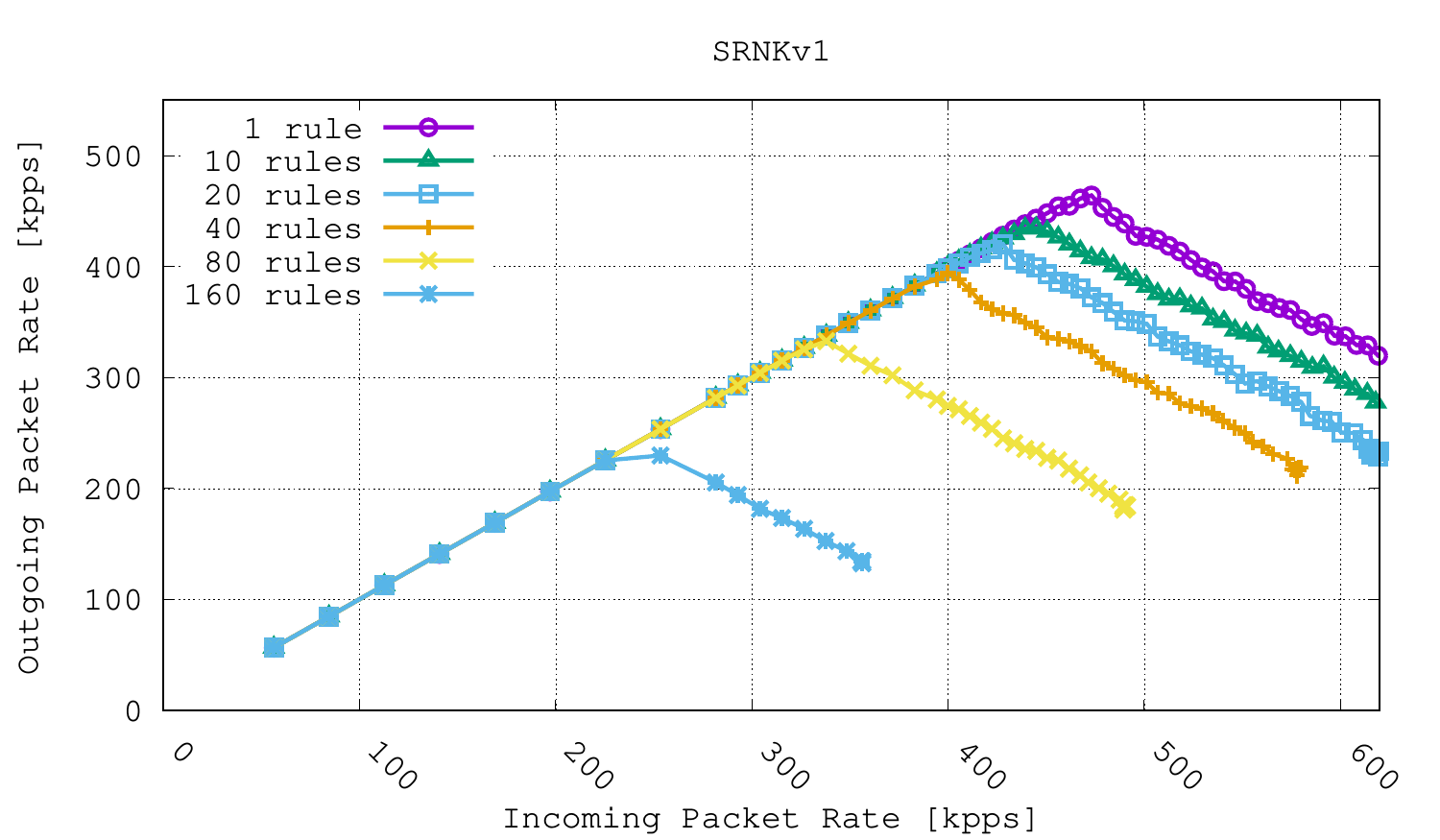}
    \caption{Throughput vs Number of rules/VNFs for the SR End.AD behaviour (SRNKv1)}
    \label{fig:testbed_3}
    \vspace{-3ex}
\end{figure}

In this appendix we report additional experimental results, shown in Fig.~\ref{fig:testbed_1}, Fig.~\ref{fig:testbed_2}, Fig.~\ref{fig:testbed_3}

%% file: inc/extended-nsh.tex
\section{NSH comparison with SRv6}
\label{sec:ext_nsh_tutorial}

\begin{figure*}[t]
    \centering
    \includegraphics[width=17.5cm, height=7.5cm]{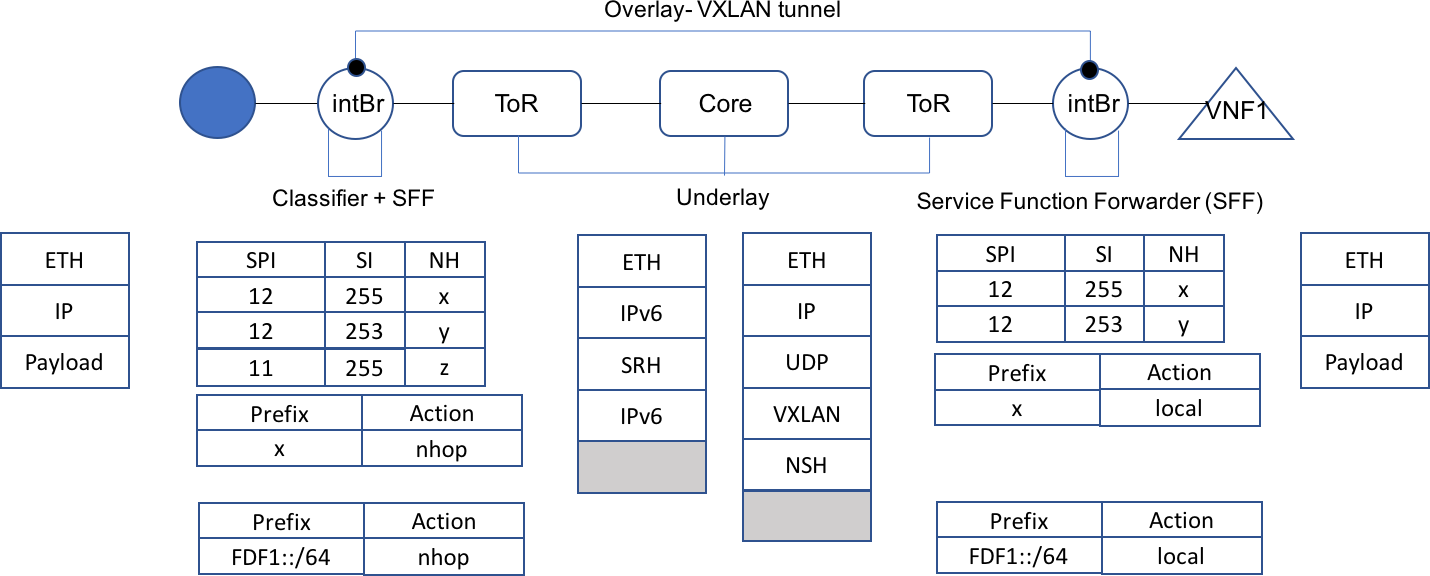}
    \caption{Comparison between NSH and SRH operations}
    \label{fig:nsh_srh}
    \vspace{-3ex}
\end{figure*}

The SRv6 solution brings a simplification of the network operations offering the possibility to implement SFC in the network without the need of having completely separated \textit{overlay} and \textit{underlay} operations. Indeed, with SRv6 these differences are ``blurred'' and we can literally program SFC in the network with several advantages. Figure \ref{fig:nsh_srh} shows a comparison between NSH and SRH solutions in terms of: i) packet encapsulation; ii) packet forwarding; iii) state to be maintained/configured in the network nodes. 

The NSH header encapsulates user traffic and then uses tunneling mechanisms to steer packets over the data center fabric. VXLAN tunnels have to be configured and maintained manually or using management protocols like ovsdb \cite{rfc7047}. On the other side traffic encapsulated with the SRH does not leverage any tunnel and can be directly forwarded by the \textit{underlay} fabric. 

Most of the times in a NSH based deployment \textit{underlay}, tunnels and SFC need to be controlled by different control planes. Usually it is necessary to implement these functions in two different network nodes: i) software switch or virtual router in the virtualization server; ii) hardware switch of the fabric. Having such as VXLAN tunnels introduces further overhead in the virtual nodes. One possible optimization would be to offload the tunneling part to the fabric but this would require the VTEP offloading functionality which could not be available on the \textit{underlay} devices. 

Network visibility would also benefit of the simplification introduced by SRv6, it would be easier to debug and troubleshoot a SRv6 based infrastructure with respect to one leveraging tunneling mechanisms and having a protocol stack of 7 or more headers. 

Finally we can state that the SRv6 based approach requires less state, NSH protocol differently from SRv6 just transports a chain identifier, the so called Service Path Identifier (SPI), and the position along the chain of the packet through the Service Index (SI) but then specific state is needed in the SFC nodes in order to proper forward the packet to the next hop. On the other side, the SRH overhead results to be bigger than the overhead introduced by NSH solutions in the worst case. 

We note that SRv6 does not necessarily need to be an alternative to NSH, indeed there are proposals that deal with the inter-working of NSH and SRv6 as envisaged in \cite{sfc_sr_nsh}. 